\documentclass[aps,prd,twocolumn,superscriptaddress,preprintnumbers,floatfix,nofootinbib,notitlepage,showkeys,showpacs]{revtex4-1}

\usepackage[utf8]{inputenc}

\usepackage{graphicx}
\usepackage{hyperref}
\usepackage{latexsym}
\usepackage{amsmath}
\usepackage{amssymb}
\usepackage{bbm}
\usepackage{ulem}
\usepackage{pdfsync}
\usepackage{epsfig}
\usepackage{epstopdf}
\usepackage{subfigure}
\usepackage{xcolor}
\usepackage{comment}
\usepackage{slashed}

\begin{document}

\title{The cosmic optical background excess, dark matter, and line-intensity mapping}

\author{José Luis Bernal}
\affiliation{William H. Miller III Department of Physics and Astronomy, Johns Hopkins University, 3400 North Charles Street, Baltimore, MD 21218, United States}

\author{Gabriela Sato-Polito}
\affiliation{William H. Miller III Department of Physics and Astronomy, Johns Hopkins University, 3400 North Charles Street, Baltimore, MD 21218, United States}

\author{Marc Kamionkowski}
\affiliation{William H. Miller III Department of Physics and Astronomy, Johns Hopkins University, 3400 North Charles Street, Baltimore, MD 21218, United States}

\begin{abstract}
Recent studies using New Horizons LORRI images have returned the most precise measurement of the cosmic optical background to date, yielding a flux that exceeds that expected from deep galaxy counts by roughly a factor of two. 
We investigate whether this excess, detected at $\sim 4\sigma$ significance, is due to axion-like dark matter that decays to 
monoenergetic photons. 
We compute the spectral energy distribution from such decays and the contribution to the flux measured by LORRI.  Assuming that axion-like particles make up all of the dark matter, the parameter space unconstrained to date that explains the measured excess spans masses and effective axion-photon couplings of 8 - 20 eV masses and 3 - 6 $\times 10^{-11}$ GeV$^{-1}$, respectively.  If the excess arises from dark-matter decay to a photon line, there will be a significant signal in forthcoming line-intensity mapping measurements that will allow the discrimination of this hypothesis from other candidates.  
\end{abstract}

\maketitle

The cosmic optical background (COB), the integrated flux of optical light, allows us to test whether our census of known galaxies is complete \cite{Cooray_review}.  Measuring the COB is challenging due to the overwhelming presence of foregrounds---especially zodiacal light (sunlight scattered by interplanetary dust)---which requires very accurate and precise calibration and modeling to be subtracted from the observations. Foreground removal can be achieved at the expense of large uncertainties in the COB measurement, which hinders detection from Earth or from Earth-orbiting satellites. There are $\sim 2\sigma$ hints of the COB, including the 0.40 $\mu$m `dark cloud' measurement~\cite{Mattila_17}, the 0.80 $\mu$m CIBER flux~\cite{Matsuura_CIBER}, and the Pioneer 10 and 11 observations~\cite{Matsuoka_pioneer} taken at $3-5$ AU from the Sun (but challenged by Ref.~\cite{Matsumoto_2018}). Nonetheless, contamination from zodiacal light is negligible in the outer solar system~\cite{Zemcov_2018,Poppe_2019}. This makes NASA's New Horizons spacecraft an extraordinary platform to measure the COB, as demonstrated by analyses using archival images of the New Horizons' Long Range Reconnaisance Imager (LORRI)~\cite{Zemcov_NHCOB, NH21}. 
The New Horizons' LORRI has recently provided, using targeted observations taken at 51.3 AU from the Sun, the first high signal-to-noise detection of the COB, yielding a flux of photons with wavelengths $\sim 0.4-0.9$ $\mu$m ($\sim 1.3-3$ eV) of $16.37\pm 1.47$ nW/m$^2$/sr \cite{NH22}. This measurement, obtained after subtracting contributions from diffuse Galactic light, scattered light from stars and galaxies outside the LORRI field, faint stars below the detection limit, hydrogen and ionized helium two-photon continua, and foregrounds from the spacecraft, exceeds the flux expected from deep Hubble Space Telescope galaxy counts by  $8.06\pm 1.92$ nW/m$^2$/sr~\cite{NH22}; this is roughly a factor-of-two excess, with $\gtrsim4\sigma$ significance. Possible astrophysical explanations include a faint population of galaxies not accounted for in the prediction from deep counts of HST~\cite{Conselice_16}, light from stars tidally removed from galaxies, a population of faint sources within extended halos (intrahalo light)~\cite{Cooray_2012, Zemcov_2014, Matsumoto_19}, or direct-collapse black holes at very high redshift~\cite{Yue_DCBH}. 

Another, more exotic, possibility is that the excess is due to dark-matter decay. Although dark matter is a cornerstone of the standard cosmological model, there is currently no satisfactory microscopic model for dark matter.  
Dark matter may decay to photons through some weak coupling to Standard Model particles; examples include the axion~\cite{Abbott:1982af, Dine:1982ah,Preskill:1982cy, Weinberg:1977ma, Wilczek:1977pj, Peccei:1977hh, Peccei:1977ur} and sterile neutrinos~\cite{Kusenko_SterNu}. There are ongoing efforts to constrain this possibility with cosmological and astrophysical observations over a vast mass range (see e.g., Refs.~\cite{Caputo:2018ljp, Caputo:2018vmy,Gong_IR,Kalashev_decDM_NIRB,Caputo_IR,Caputo:2019djj, Boyarsky:2014ska, Riemer-Sorensen:2014yda, Dessert:2018qih,Blanco_gammaray,Cohen_gammaray,NathMaity_Pevdm,Esmaili_pevdm,Ellis_gravitino, Hu_DMSpecD, Chluba_specD,Iocco_BBN,Pospelov_BBN,Poulin_BBNDM,Poulin_decDM,Slatyer_dark,Slatyer_decDM,Lucca_synergy}). 

Here we show that if the COB excess is due to dark-matter decay to a monoenergetic photon, this decay line will be detected with high significance with forthcoming line-intensity mapping (LIM) experiments \cite{SPHEREx,HETDEX,Kovetz:2017agg,Bernal:2019jdo}.  LIM experiments infer the three-dimensional distribution of mass by mapping the cosmic luminosity density of some specific atomic/molecular emission line; e.g., the 21-cm transition of neutral hydrogen, rotational lines of carbon monoxide, a bright infrared line of singly-ionized carbon, or the Lyman- and Balmer-series of neutral hydrogen.  If dark matter decays to a photon line, it will show up in LIM experiments as an unidentified line \cite{Creque-Sarbinowski:2018ebl,Bernal_limdm}, allowing the discrimination between potential candidates to explain the COB excess.  

To be concrete, we focus on the two-photon decay of an axion that makes up all the dark matter. For an axion mass $m_\chi$, the decay rate $\Gamma_\chi$ is related to the effective coupling  $g_{\chi\gamma}$ to photons as $\Gamma_\chi = (m_\chi c^2)^3g_{\chi\gamma}^2/32h$, where $c$ is the speed of light and $h$ is the Planck constant. In this case, the energy density in the COB excess, which is $\sim 10^{-6}$ of the dark-matter density, implies an axion lifetime $\sim 10^6$ times the age $t_{\rm U}$ of the Universe, and so we assume the decay lifetime is long compared with $t_{\rm U}$ and neglect any variation of the dark matter density with time due to decays. The photons produced in the axion decay have a rest-frame frequency $\nu=m_\chi c^2/2h$. In the context of the QCD axion, the  mass range corresponding to the LORRI band requires couplings that are ruled out by observations~\cite{Adams:2022pbo}. Nonetheless, similar production processes may operate in the early Universe to produce so-called axion-like particles (ALPs), which might be the dark matter and do not have strict relations between their mass and the coupling to photons~\cite{Arias:2012az}. As we show, ALP decays contributing to explain the COB excess are fully consistent with all existing constraints to the axion parameter space.  

The profile of the emission coming from a given redshift has a width dependent on the dark matter velocity dispersion, which is negligible for the frequencies of interest~\cite{Gong_IR}; hence, we consider a Dirac delta function profile. The specific intensity $I_\lambda$ per observed wavelength $\lambda_{\rm obs}$ in units of power per area per steradian for the dark matter radiative decay is given by
\begin{equation}
    I_\lambda = \frac{c}{4\pi}\frac{f(z)\Omega_{\rm dm}\rho_c  c^2\Gamma_\chi}{\lambda_{\rm obs}(1+z)H(z)}\,,
    \label{eq:Ilambda}
\end{equation}
where $\Omega_{\rm dm}$ and $\rho_c$ are the dark matter density parameter and the critical density today, respectively, $\Gamma_\chi$ is the dark matter decay rate, and $H$ is the Hubble expansion rate at the redshift $z\equiv z(m_\chi,\lambda_{\rm obs})$ at which the decay occurs, which in turns depends on the dark matter mass and the observed wavelength. Assuming all ALPs have the same mass, the spectral energy distribution at each observed wavelength only receives contribution from a single redshift. The numerator of the second factor is the mean specific luminosity density per comoving volume produced by dark matter decays~\cite{Chen:2003gz, Finkbeiner:2011dx, Pierpaoli:2003rz}, to which we add a fraction $f$ --- that may include redshift dependence --- accounting for the possibility that not all the dark matter decays, that it may decay to particles other than photons, and that not all the photons may reach us; hereinafter we assume this fraction to be unity, but our results can be reinterpreted by linearly rescaling $\Gamma_\chi$ to account for any combination of these scenarios. We have also neglected the contribution from stimulated decays due to background light~\cite{Caputo:2018ljp,Caputo:2018vmy}, since it does not affect the result for the energies considered in this work~\cite{Bernal_limdm}. We transform the luminosity density into flux (volume) density dividing by $4\pi D_L^2$, where $D_L$ is the luminosity distance, and convert it to specific intensity by transforming the comoving volume element to solid angle and observed wavelength elements as ${\rm d}A/{\rm d}\Omega=D_M^2$ and ${\rm d}\chi/{\rm d}\lambda_{\rm obs}=c(1+z)/H\lambda_{\rm obs}$, respectively, where $D_M$ and $\chi$ are the comoving angular diameter distance and the comoving radial distance, respectively. 

\begin{figure}[t]
 \begin{centering}
\includegraphics[width=\columnwidth]{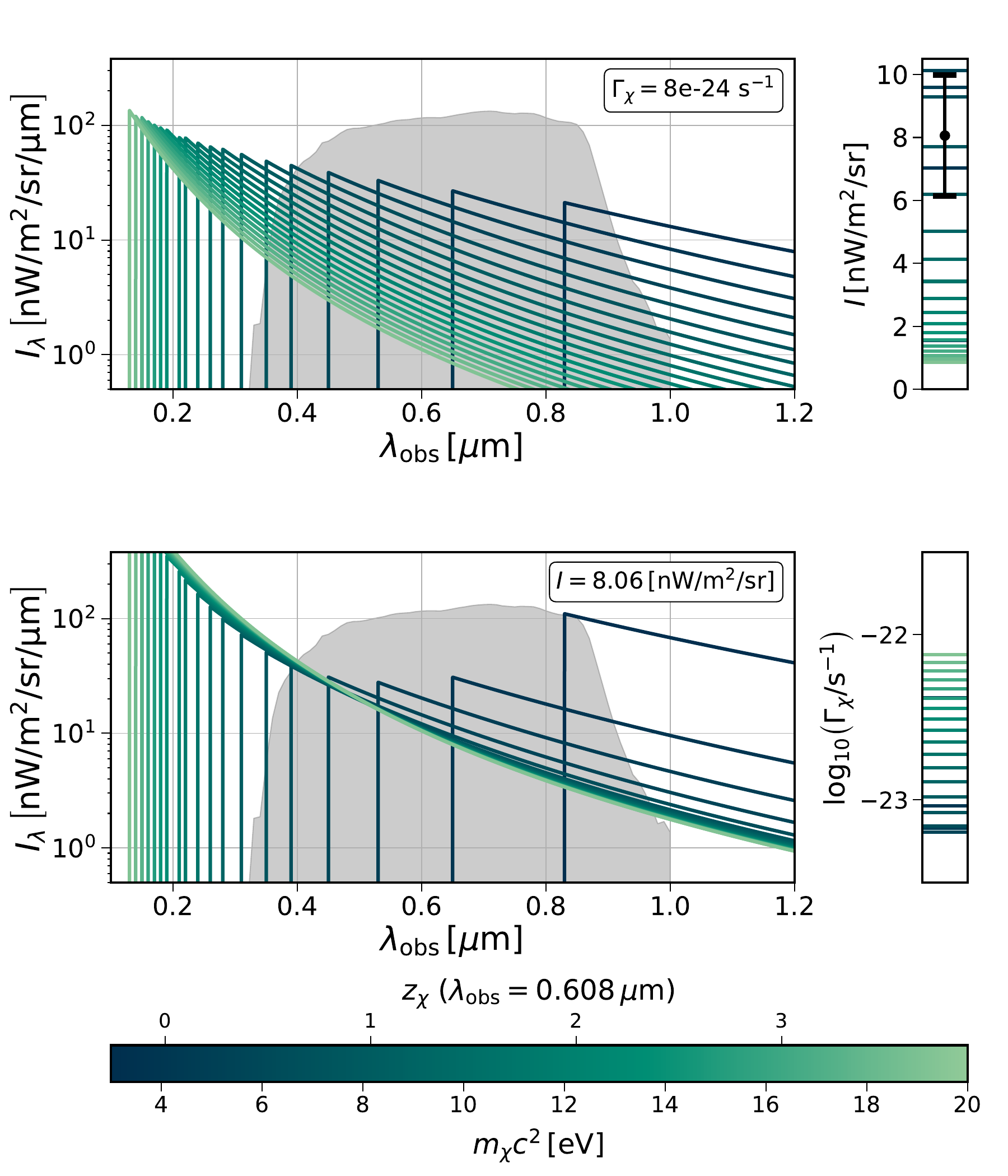}
\caption{Specific intensity per observed wavelength from  ALP radiative decays 
as ALP mass varies (color coded following the colorbar). We show predictions for a fixed decay rate of $8\times 10^{-24}$ s$^{-1}$ (top left panel), and for a fixed total contribution to the measured intensity equal to the COB excess (bottom left panel), with the corresponding contributions to measured intensities along with the COB excess (top right panel) and required decay rates to match the COB excess (bottom right panel).  
We also indicate in the colorbar the redshift at which the decay occurs at the LORRI pivot wavelength $\lambda_{\rm obs}=0.608\,\mu$m for each mass. The gray shaded region corresponds to the unnormalized shape of the LORRI responsivity as function of wavelength.}  
\label{fig:spectrum}
\end{centering}
\end{figure}

We show the specific intensity per observed wavelength for different  ALP masses over the interval of interest in Fig.~\ref{fig:spectrum}. The hard cut off at the low wavelength end of each curve corresponds to the rest-frame wavelength of the products for each ALP mass, i.e., to the radiation at $z=0$. For a fixed decay rate, the flux decays with the ALP mass, which also implies that it decays as the radiation comes from higher redshift at the same $\lambda_{\rm obs}$.This is due to the fact that the flux density decreases with distance: accounting for all redshift-dependent factors, $I_\lambda\propto (1+z)^{-5/2}$ (roughly approximating $H\propto (1+z)^{3/2}$). 

LORRI is an unfiltered CCD imager with sensitivity from the blue to the near infrared --- $0.4\,\mu{\rm m}\lesssim \lambda_{\rm obs}\lesssim 0.9\,\mu{\rm m}$ --- with pivot wavelength $\lambda_{\rm piv}=0.608\,\mu{\rm m}$, which is operated with $4\times 4$ pixel binning for deep observations~\cite{Cheng_LORRI,LORRI}. We use Eq.~\eqref{eq:Ilambda} and the specifications of the LORRI camera in its $4\times 4$ operating mode to obtain the predicted contribution from dark matter decays to the COB measurement. We compute the data number DN per second and pixel using the responsivity $R$ of LORRI~\cite{LORRI} as
\begin{equation}
   {\rm DN}/{\rm pix}/{\rm s} = \int {\rm d}\lambda_{\rm obs}R(\lambda_{\rm obs})I_\lambda\,,
\end{equation}
which we convert into measured intensity at $\lambda_{\rm piv}$ using the sensitivity corresponding to the diffuse photometry keyword for targets with Solar-like spectral energy distributions (RSOLAR in Table 2 of Ref.~\cite{LORRI}, transformed to the correct units):
\begin{equation}
    I = \frac{\lambda_{\rm piv}}{{\rm RSOLAR}} \times\left({\rm DN}/{\rm pix}/{\rm s}\right)\,.
\label{eq:I}    
\end{equation}
We show the predicted contributions to the measured flux by LORRI in the right panel of Fig.~\ref{fig:spectrum}, along with the  COB excess measured, which could be explained by the correct combination of ALP mass and decay rate. In the bottom panels we show the specific intensities for each mass normalized to yield the correct contribution to the measured flux to explain the COB excess, indicating the corresponding decay rate in the right panel.

\begin{figure}[t]
 \begin{centering}
\includegraphics[width=\columnwidth]{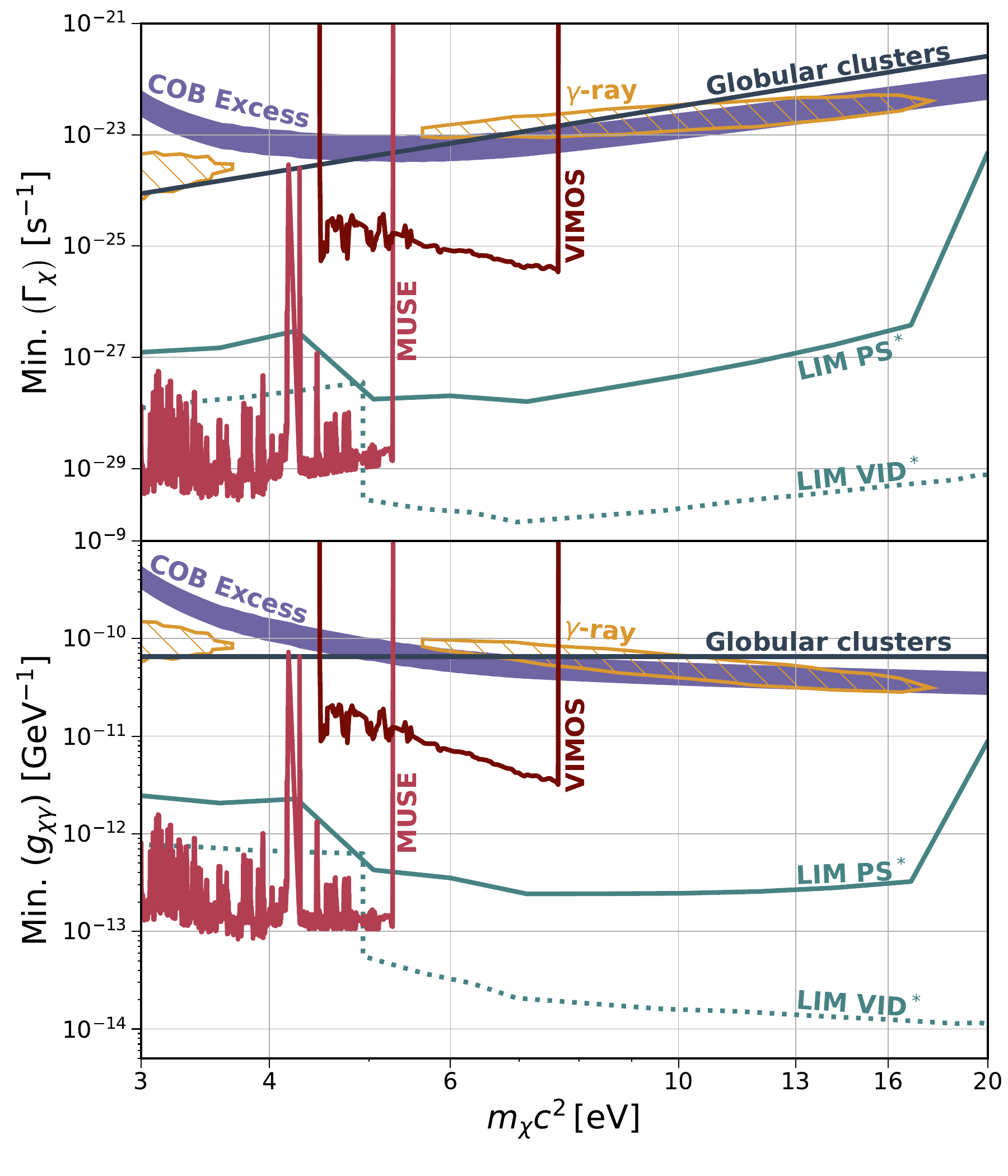}
\caption{Comparison between the required decay rates (top panel) and the corresponding effective ALP-photon coupling (bottom panel) to explain the cosmic optical background and existing bounds 
from spectroscopic searches and the cooling of horizontal branch stars in globular clusters, as well as forecasted sensitivities for optical LIM experiments (solid and dotted lines show the forecasts using the LIM power spectrum and voxel intensity distribution, respectively). All values correspond to 95\% confidence level, except for the 68\% confidence level favored regions from the analysis of $\gamma$-ray attenuation and claimed HST limits.}
\label{fig:constraints}
\end{centering}
\end{figure}

We calculate the required decay rates to explain the COB excess as function of mass from Eq.~\eqref{eq:I}. We limit our study to $3\,{\rm eV}\lesssim m_\chi c^2\lesssim 20\,{\rm eV}$, since photons produced from more massive ALP are heavily absorbed by intergalactic gas along the line of sight (see e.g.,~\cite{1995ApJ...441...18M, Inoue:2014zna}.) In addition, that range of the parameter space is further constrained by alternative observations~\cite{Wadekar:2021qae, Bolliet:2020ofj}. The required decay rates (and corresponding photon-axion couplings) at 95\% confidence level are shown in Fig.~\ref{fig:constraints}, along with existing constraints and forecasted sensitivities. We include the most competitive existing bounds on this mass range, namely the spectroscopic observations of the dwarf spheroidal galaxy Leo T with MUSE~\cite{Regis:2020fhw} and optical searches using VIMOS spectra for line emission in the galaxy clusters Abell 2667 and 2390~\cite{Grin:2006aw}, and limits derived from the study of the cooling of horizontal branch stars in globular clusters~\cite{Ayala:2014pea}. We also include the tentative 68\% confidence level favored regions on the axion parameter space from the study of the blazar 1ES 1218+304 $\gamma$-ray spectrum~\cite{Korochkin_axion}, which requires a bump in the extragalactic  background light (EBL) to fit the attenuation of the spectrum at high energies~\cite{Korochkin_bump}. For $8\, {\rm eV}\lesssim m_\chi c^2 \lesssim 20\, {\rm eV}$ the required decay rates are currently unconstrained, and interestingly overlap with the favored region from Ref.~\cite{Korochkin_axion}. We do not include existing constraints from Ref.~\cite{Cadamuro:2011fd} using observations of the optical and near-infrared background, since the substantial update in this kind of observations (including the COB excess discussed in this paper) requires the revision of those bounds.

Thus, it is possible that the COB excess is indeed produced by 
ALP decays. Additional, independent observations will be required to confirm or reject this candidate. Radiative decays of axion-like particles have been considered to explain the anomalous excesses in the power spectrum of the  near-infrared background measured by HST~\cite{HST_NIRPS}, CIBER~\cite{Zemcov_NIRPS} and \textit{Spitzer}~\cite{Cooray_2012,Kashlinsky_spitzer}. The predicted axion parameters that provide a good fit to the measurements~\cite{Gong_IR,Caputo_IR} roughly coincide with those that would explain the COB excess in the low-mass end of the mass range considered in this work; however, this part of the parameter space is ruled out by independent probes~\cite{Regis:2020fhw,Ayala:2014pea}.

The region of the parameter space of axion properties that could explain the COB excess has proved difficult to explore with current strategies. The specific intensities for some of the masses required to explain the COB excess also predict contributions in the ultraviolet (UV) range (see Fig.~\ref{fig:spectrum}). New Horizons also includes a UV spectrograph that, if pointed to dark fields as the LORRI camera was, could measure the cosmic UV background and shed light on whether the COB excess is produced by dark matter decays. 

Here, however, we emphasize the prospects to test this 
ALP decay scenario with forthcoming LIM experiments.  The principle underlying LIM searches for radiative dark matter decays is similar to that for a background excess, with the exception of the spectral resolution embedded in LIM experiments.\footnote{Since LIM experiments target spectral lines that can be well distinguished from the continuum, their sensitivity lowers as the emission line or the dark matter mass distribution widens.} In this case, the emission line produced via dark-matter decay redshifts into the telescope spectral band from a different redshift than the targeted astrophysical lines, acting as what is known as a line interloper. The dark matter line can therefore be detected using standard data analysis techniques (see e.g., Refs.~\cite{Lidz_interlopers, Cheng_foregroundsAP, Cheng_deconfInterlopers, Gong_interlopers}). The most prominent features decaying dark matter imparts on LIM observables are an increase in the power spectrum anisotropy due to projection effects and a voxel intensity distribution that is narrower and shifted towards higher temperatures. In particular, LIM experiments in the optical such as HETDEX~\cite{HETDEX} and SPHEREx~\cite{SPHEREx} will improve current sensitivities by several orders of magnitude, being able to detect the contributions from dark matter decays required to explain the COB excess with very high significance. We show the forecasted 95\% combined sensitivities~\cite{Bernal_limdm} from LIM power spectrum (solid) and voxel intensity distribution (dotted) measurements by these experiments; the measurements of these two observables can be combined to further improve these sensitivities~\cite{COMAP:2018kem,SatoPolito_covariance}. The  dark matter line can also be isolated by cross-correlating the line intensity maps with galaxy surveys~\cite{Creque-Sarbinowski:2018ebl,Shirasaki_limlensdm}. Furthermore, LIM will be able to distinguish between different exotic line emitters through the clustering of the fluctuations of the associated intensity~\cite{Bernal:2021ylz}. In turn, a non-detection of exotic line emission is detected in LIM observations would imply that the COB excess is produced by sources with extended spectra energy distributions.

Current measurements of blazar high-energy spectrum attenuation via photon-photon electron-positron pair production~\cite{HESS, Ahnen16, Fermilat} can be used to reconstruct the EBL at ultraviolet, optical and infrared wavelengths as function of redshift. When a tighter parametrization of this flux is assumed, the reconstructed background emission~\cite{Ahnen16,Fermilat,Desai19} agrees with the predicted integrated galaxy light from galaxy counts~\cite{NH21,Driver_IGL,SaldanaLopez_IGL}. The uncertainties, once more freedom is given to the shape of the reconstructed specific intensity of the background radiation~\cite{HESS_background,Acciari19}, are large enough to agree with predictions from number counts and the measured COB excess. As discussed in Refs.~\cite{Kalashev_decDM_NIRB,Korochkin_axion}, a higher COB from high-redshift sources will contribute to the $\gamma$-ray attenuation, which can be used to look for axion decays if secondary $\gamma$-ray photons created from ultra-high energy cosmic rays emitted by the blazar and that are not deflected by magnetic fields are neglected~\cite{Essey:2009ju, Essey:2009zg, Essey:2010er}. Preliminary results from the study of the blazar 1ES 1218+304~\cite{Korochkin_axion} show preference for a bump in the EBL. This result, coherent with the LORRI measurements, adds significance to the claim for a general EBL excess. Furthermore, the fact that both excesses may be contributed by ALPs with similar properties make the ALP interpretation more coherent. This hint, in addition to the promising prospects of forthcoming high-energy observations, warrants further investigation~\cite{axion_gray_constraint}. 

We have considered ALP radiative decays, but photons can be produced from particle cascades of other decay products; 
in that case the mapping between their energy and the dark matter mass  is less direct. This scenario can be accounted for by considering a given dark matter model and modeling the final photon energy distribution and branching ratio for each decay channel; hence, our results and calculations can be extrapolated to each specific case. Similarly, dark matter annihilation includes the dark matter velocity distribution and boost factor, which adds uncertainty to the computation. Nonetheless, only fermionic dark matter can involve efficient annihilation, and dark matter with masses below $\sim 100$ eV can only be bosonic due to the Pauli exclusion principle. Thus, if we consider that dark matter is a single species, it cannot annihilate producing photons in the frequency band of interest. We leave the study of a small fraction of light fermionic annihilating dark matter for future work. Another dark-matter related potential explanation for the COB excess involves dark-matter powered stars~\cite{Maurer_DS}.

The results of this work provide a potential explanation for the COB excess that is allowed by independent observational constraints and that may answer one of the most long-standing unknown in cosmology: the nature of dark matter. We provide additional motivation for follow up studies with forthcoming observations, especially with optical LIM surveys such as HETDEX or SPHEREx, with the UV instrument aboard the New Horizons spacecraft, and very high-energy $\gamma$-ray attenuation, that will confirm or reject this hypothesis.

\hfill

\textit{Acknowledgments} ---
The authors thank Marc Postman for help with the LORRI responsivity and useful discussions. JLB is supported by the Allan C.\ and Dorothy H. Davis Fellowship. GSP was supported by the National Science Foundation Graduate Research Fellowship under Grant No.\ DGE1746891.  MK was supported by NSF Grant No.\ 1818899 and the Simons Foundation.

\bibliography{ref.bib}
\bibliographystyle{utcaps}
\end{document}